\documentclass[aps,pre,twocolumn,a4paper,10pt,notitlepage,showpacs,nofootinbib,superscriptaddress]{revtex4-1}%
\usepackage[english]{babel}
\usepackage[T1]{fontenc}
\usepackage{endnotes}
\usepackage{amssymb,amsmath,amsfonts}
\usepackage{textcomp}
\usepackage{graphicx,color}

\newcommand{\dif}{\mathrm{d}}%
\newcommand{\fdif}{\operatorname{\delta}}
\newcommand{\Fdif}[2]{\frac{\fdif\!#1}{\fdif\!#2}}

\renewcommand{\vec}[1]{\boldsymbol{\mathrm{#1}}}%

\begin{document}

\title{Active Model H: Scalar Active Matter in a Momentum-Conserving Fluid}

\author{Adriano Tiribocchi}
%\email{}
\affiliation{SUPA, School of Physics and Astronomy, University of Edinburgh, Edinburgh EH9 3FD, United Kingdom}
\affiliation{Department of Physics and Astronomy, University of Padua, I-35131 Padova, Italy}

\author{Raphael Wittkowski}
\affiliation{SUPA, School of Physics and Astronomy, University of Edinburgh, Edinburgh EH9 3FD, United Kingdom}
\affiliation{Institut f\"ur Theoretische Physik II,  Heinrich-Heine-Universit\"at D\"usseldorf, D-40225 D\"usseldorf, Germany}

\author{Davide Marenduzzo}
\affiliation{SUPA, School of Physics and Astronomy, University of Edinburgh, Edinburgh EH9 3FD, United Kingdom}

\author{Michael E. Cates}
\affiliation{SUPA, School of Physics and Astronomy, University of Edinburgh, Edinburgh EH9 3FD, United Kingdom}

\date{\today}

\begin{abstract}
We present a continuum theory of self-propelled particles, without alignment interactions, in a momentum-conserving solvent. 
To address phase separation we introduce a scalar concentration field $\phi$ with advective-diffusive dynamics. Activity creates a contribution $\Sigma_{ij}=-\zeta((\partial_i\phi)(\partial_j\phi)-(\nabla\phi)^{2}\delta_{ij}/d)$ to the deviatoric stress, where $\zeta$ is odd under time reversal and $d$ is the number of spatial dimensions; 
this causes an effective interfacial tension contribution that is negative for contractile swimmers. 
We predict that domain growth then ceases at a length scale where diffusive coarsening is balanced by active stretching of interfaces, and confirm this numerically. 
Thus the interplay of activity and hydrodynamics is highly nontrivial, even without alignment interactions.
\end{abstract}


\maketitle

`Active matter' means the collective dynamics of self-propelled particles at high density. By converting energy into motion, such particles violate time-reversal symmetry (TRS) at the micro-scale. This violation changes the structure of coarse-grained equations of motion, allowing far-from-equilibrium physics to dominate at large scales \cite{RMP}. Active matter can be `wet', i.e., coupled in bulk to a momentum-conserving solvent, or `dry', i.e., for instance in contact with a momentum-absorbing wall.\footnote{This wet/dry terminology has become conventional, although many `dry' systems are immersed in a fluid.} `Wet' active systems include not only bacterial swarms in a fluid, the cytoskeleton of living cells, and biomimetic cell extracts \cite{RMP,Cisneros,Actomyosin,Dogic}, but also synthetic colloidal swimmers in a fully bulk geometry. Such swimmers may in future offer a toolbox for directed assembly of new materials in three dimensions. Many of these artificial colloidal swimmers are spherical objects, with asymmetric coatings that cause them to move through a bath of fuel and/or under illumination with light \cite{synthetics}.  
 
Of particular importance is the `active liquid crystal' (ALC) theory \cite{RMP,SRreview}, which starts from a passive fluid of rod-like objects \cite{BerisEdwards} with either a polar order parameter $\vec{P}$ \cite{Joanny}, or a nematic one $\mathbf{Q}$ \cite{Simha}. An active stress is then added; this is $-\zeta \vec{P}\!\otimes\!\vec{P}$ or $-\zeta \mathbf{Q}$, with $\zeta$ odd under time reversal and the dyadic product $\otimes$, representing the leading-order TRS violation in an orientationally ordered medium. This causes new physics such as giant number fluctuations \cite{GNF}, and spontaneous flow above an activity threshold that vanishes for large systems \cite{Joanny,Simha}. This instability depends on whether particles are extensile (pulling fluid inwards equatorially and emitting it axially) or contractile (vice versa). Numerical solutions \cite{MarenduPRE,CristinaDefect,Thampi} show spontaneous flows resembling experiments on bacterial swarms \cite{Cisneros} and on microtubule-based cell extracts \cite{Dogic}. 

There is one important effect of activity that ALC models do not capture (unless added by hand \cite{Farrell}): motility-induced phase separation (MIPS) \cite{TC,CT}. If their propulsion speed falls fast enough with density (e.g., due to crowding interactions), even purely repulsive active particles phase-separate into dense and dilute domains. MIPS is by now well-established, at least for `dry' models such as Brownian dynamics simulations of self-propelled colloidal spheres \cite{fily,baskaran,sten1,sten2}. Unlike the physics of ALCs, MIPS does not depend on alignment interactions \cite{fily}, and can be captured at continuum level by a scalar concentration field $\phi$ alone, without direct appeal to order parameters $\vec{P}$ or $\mathbf{Q}$ \cite{sten1,NatCom}. This is pertinent to spherical colloidal swimmers \cite{synthetics}, which (although not entirely devoid of alignment interactions \cite{SMF}) are not orientationally ordered in the passive limit, contrary to the ALC premise. The theory of ``scalar active matter", with MIPS as its main feature, has been developed so far only for dry systems \cite{TC,CT,MicrobeReview,TCReview}. Following the same path as for passive systems \cite{HH,Bray} culminates in a dynamical scalar $\phi^4$ field theory, called `Active Model B' \cite{NatCom}.  

\begin{figure}[ht]
\begin{center}%
\includegraphics[width=\linewidth]{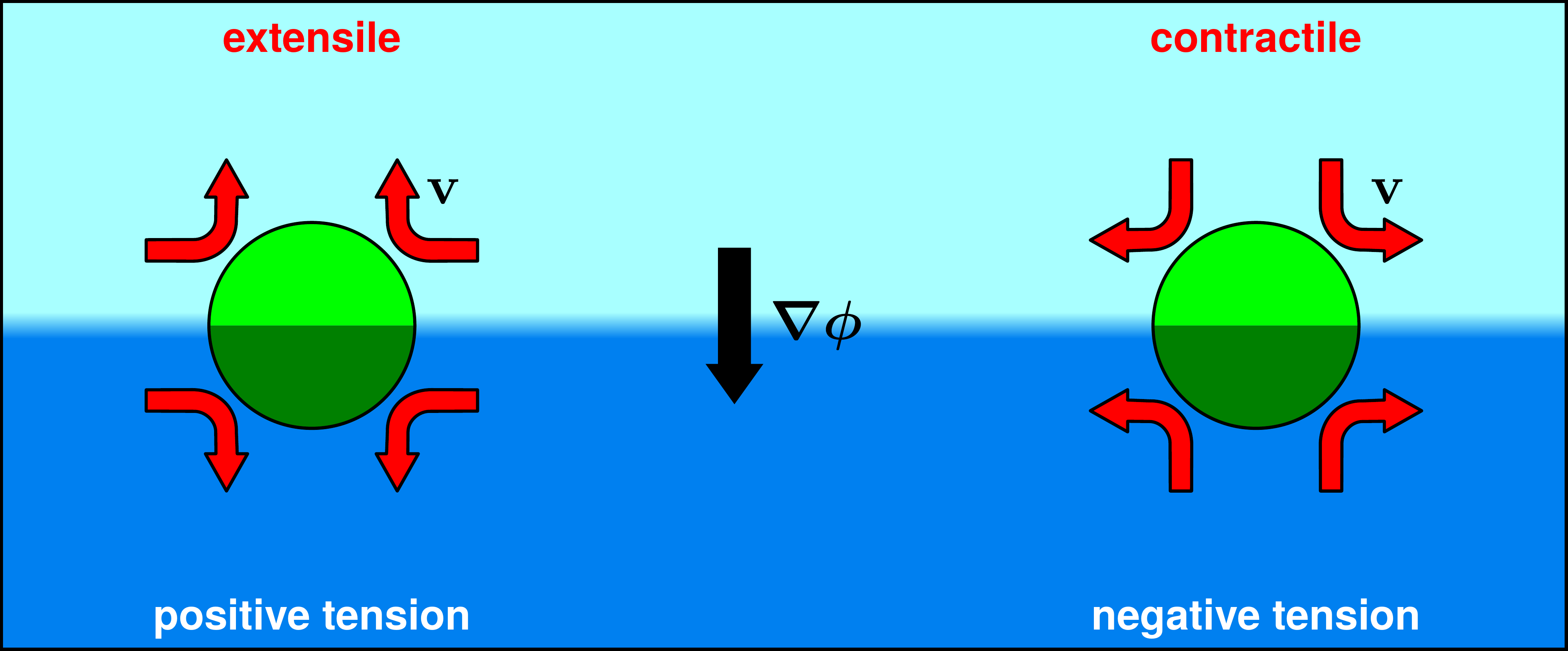}\vspace*{-1ex}%
\caption{Schematic diagram showing the flow caused by swimmers whose polarization is normal to an interface between phases. For contractile swimmers this is mechanically equivalent to a negative interfacial tension.}
\label{fig1}%
\end{center}%
\end{figure}

In this Letter we extend this simplest scalar active matter model to the wet case. For passive systems it is well known how to couple a diffusive, conserved, phase-separating order parameter (see passive Model B \cite{HH}) to an isothermal fluid flow; the result is called `Model H' \cite{HH}. We follow a similar path, but find that activity deeply alters the relation between the stress and the scalar order parameter. At first sight, our final equations resemble closely those of passive Model H, in which the domain size $L(t)$ scales linearly in time, $L\propto \sigma t/\eta$, as found from a force-balance between interfacial tension $\sigma$ and dynamic viscosity $\eta$. On closer inspection though, the active equations involve two separate tensions, one in the diffusive and one in the mechanical sector. The second of these can be (and for pure MIPS actually is) \textit{negative for contractile particles}. As a result, while extensile systems show relatively normal domain growth, contractile ones should cease to coarsen at a certain  scale, set by a balance between loss of interfacial area through diffusive coarsening, and its creation under the action of the contractile stress. The latter effect arises because swimming particles are more likely to point normal to a fluid-fluid interface than tangential to it (see Fig.\ \ref{fig1}). Within a MIPS context, we also show that this orientational bias has a purely kinematic explanation: it does not require interparticle torques.

\textit{Active Model B:} To establish notation and basic concepts, we briefly recapitulate the formulation of \cite{NatCom}. For `dry' scalar active matter, the diffusive dynamics of the compositional order-parameter field $\phi(\vec{r},t)$ obeys\footnote{In \cite{NatCom} it was shown that Eqs.\ \eqref{one}-\eqref{three} capture all terms in $\vec{J}$ to order $\nabla^3$ and $\phi^2$, with one exception, $\epsilon\phi\nabla^3\phi$, which can be removed by introducing a density-dependent mobility; see Appendix \ref{AI}.}
\begin{gather}
\dot\phi = -\nabla\!\cdot\!\vec{J} \;, \label{one}\\
\vec{J} = -\nabla \mu + \vec{\Lambda} \;, \label{two}\\
\mu = a\phi + b\phi^3 -\kappa \nabla^2\phi + \lambda (\nabla\phi)^2 \;. \label{three}
\end{gather}
Here, $\vec{J}$ is the diffusive current, whose mobility $M$ in Eq.\ \eqref{two} is set to unity, $\vec{\Lambda}$ is a zero-mean, unit-variance Gaussian white noise, and $a$, $b$, $\kappa$, and $\lambda$ are constants. (Negative $a$ is chosen to ensure phase separation; $b$ and $\kappa$ are positive for stability.) 
This equation set differs from the traditional diffusive dynamics of a conserved scalar field (see Model B \cite{HH}) solely by addition of a leading-order TRS violation, namely the $\lambda$ term. Without this term, $\mu$, which resembles a chemical potential, can be written as $\delta\mathcal{F}/\delta \phi$ with the functional (setting $k_BT=1$)
\begin{equation}
\mathcal{F} = \int\!\left(\frac{a}{2}\phi^2+\frac{b}{4}\phi^4 +\frac{\kappa}{2}(\nabla\phi)^2\right) \dif^{d}r \;, 
\label{four}% 
\end{equation}
where $d$ is the number of spatial dimensions. 
For active systems $\mathcal{F}$ is not a genuine free-energy functional, and in the simplest MIPS theory, it stems solely from the density-dependence of the propulsion speed \cite{TC}. Nonetheless, if $\lambda$ happens to vanish, Eqs.\ \eqref{one}-\eqref{three} coincide with Model B,
which describes a passive system with free energy $\mathcal{F}$ \cite{HH}. 
In contrast, for nonzero $\lambda$ no functional $\mathcal{F}$ exists for which $\mu = \delta\mathcal{F}/\delta\phi$ \cite{NatCom}, giving Active Model B. Thus only the $\lambda$ term violates TRS
in the macroscopic equations, even if the physical origins of $a$, $b$, and $\kappa$ also do so microscopically. 

\textit{Active Model H:} We now wish to couple the diffusive dynamics of $\phi(\vec{r},t)$ to a momentum-conserving solvent with fluid velocity $\vec{v}(\vec{r},t)$. Diffusive dynamics now takes place in the frame of the moving fluid so that Eq.\ \eqref{one} acquires an advective time derivative, 
\begin{eqnarray}
\dot\phi+\vec{v}\!\cdot\!\nabla\phi &=& -\nabla\!\cdot\!\vec{J} \;, \label{five}
\end{eqnarray}
with no change to Eqs.\ \eqref{two} and \eqref{three}. The fluid is incompressible,
\begin{equation}
\nabla\!\cdot\!\vec{v} = 0 \;, 
\label{six}
\end{equation}
and of unit mass density. The Navier-Stokes equation for momentum conservation then reads
\begin{equation}
\dot{\vec{v}} + \vec{v}\!\cdot\!\nabla\vec{v} = \eta\nabla^2\vec{v} - \nabla p + \nabla\!\cdot\!\mathbf{\Sigma} \;. \label{seven}
\end{equation}
Here, the pressure field $p(\vec{r},t)$ subsumes \textit{all} isotropic stress contributions and enforces Eq.\ \eqref{six}. The deviatoric stress $\mathbf{\Sigma}$ is traceless and (without orienting interactions) symmetric. 

In passive systems, $\mathbf{\Sigma}$ can be derived from the free-energy functional $\mathcal{F}$ by standard procedures \cite{Chaikin}; restoring isotropic terms one finds $\nabla\!\cdot\!\mathbf{\Sigma} = -\phi\nabla\mu$ which is the thermodynamic force density arising from concentration gradients \cite{Bray}. The deviatoric stress can then be written 
\begin{equation}
\Sigma_{ij}= -\zeta\big((\partial_i\phi)(\partial_j\phi)-\tfrac{1}{d}(\nabla\phi)^{2}\delta_{ij}\big)
\label{eight}
\end{equation}
with $\zeta = \kappa$. But if we relax that equality, Eq.\ \eqref{eight} remains the \textit{only} deviatoric stress that can be created from $\phi(\vec{r},t)$ to order $\mathcal{O}(\phi^2,\nabla^3)$,\footnote{This can be established by constructing all traceless and symmetric second-rank tensors $\mathbf{T}$ to that order, and showing that each of their divergences varies as $\nabla\!\cdot\!\mathbf{\Sigma}$ up to the gradient of a scalar field (i.e., a pressure gradient); see Appendix \ref{AII}.} and is hence the sole leading-order deviatoric stress contribution for scalar active matter.\footnote{The parameter $\zeta=\zeta_R+\zeta_D$ contains now a reversible contribution $\zeta_R$ (even under time reversal) and a dissipative one $\zeta_D$ (odd under time reversal) that stems solely from activity. If linear irreversible thermodynamics holds, we have $\zeta_R=\kappa$. This is, however, not the case for systems that are sufficiently far from equilibrium \cite{Brand}.} This neither contradicts, nor depends on, recent analyses of the pressure and/or stress in specific active models \cite{Yang,Brady,Solon1,Solon2,Bialke}.

Our Active Model H comprises Eqs.\ \eqref{two}, \eqref{three}, and \eqref{five}-\eqref{eight}. Regardless of the underlying cause of phase separation (conventional attractions and/or MIPS in any combination), only
two activity-dependent parameters distinguish the dynamics of Active Model H from a passive Model H with the free-energy functional $\mathcal{F}$ given in Eq.\ \eqref{four}, namely $\lambda$ and $\zeta$.

Our previous work on Active Model B shows $\lambda$ to have benign effects on dynamics: its main effect is to shift the coexistence condition between phases which now obey an `uncommon tangent' construction \cite{NatCom,sten1,sten2}. Once this shift is allowed for, standard diffusive dynamics (e.g., Ostwald ripening \cite{Bray}) holds qualitatively; although simulations show deviations from the expected power law $L\propto t^{1/3}$, these likely represent a slow crossover to that asymptote \cite{NatCom}. Below we thus focus mainly on the role of $\zeta$.

\begin{figure}[ht]
\centering
\includegraphics[width=\linewidth]{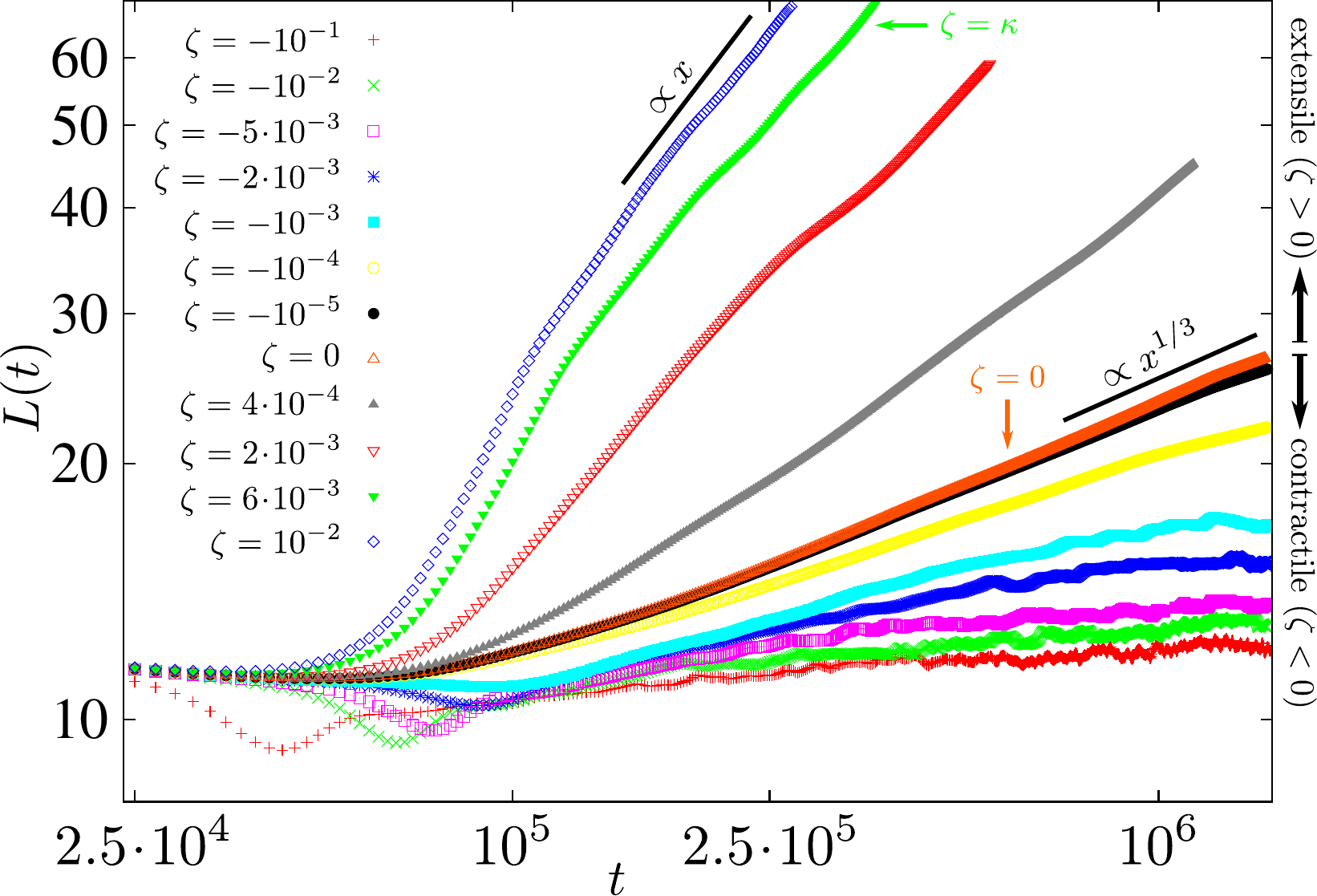}\vspace*{-1ex}%
\caption{Results for the domain size $L(t)$ in extensile ($\zeta>0$) and contractile ($\zeta<0$) two-dimensional scalar active fluids.}
\label{fig2}%
\end{figure}

\textit{Physics of the stress term:} This can be addressed in two contrasting limits. Suppose first we have a system with strong interparticle attractions, which would phase separate without activity. 
Then the thermodynamic, zeroth-order result is $\zeta = \kappa$ as mentioned previously. Perturbative activity will change this only slightly. TRS is broken, but as with $\lambda$ the effects on phase ordering dynamics should be benign. Specifically, we can use standard power-counting arguments \cite{Bray,Kendon} to estimate the dynamics of the domain size $L(t)$. 
Setting $\lambda = 0$ for simplicity and ignoring inertia (which leads to a third regime at very late times \cite{Furukawa,Kendon}), we find that (see Appendix \ref{AIII})
\begin{equation}
\dot L \simeq \alpha \sigma/L^2 + \beta \tilde\sigma/\eta \;. 
\label{nine}%
\end{equation}
Here, $\alpha$ and $\beta$ are dimensionless constants; $\sigma = (-8\kappa a^3/9b^2)^{1/2}$ is the interfacial tension in $\phi^4$ theory. When activity is nonzero but small, $\zeta \neq \kappa$ and the tension $\tilde\sigma$ differs slightly from $\sigma$. The two contributions to $\dot L$ in Eq.\ \eqref{nine} are related to gradients in $\mu$ and thus Laplace pressure $\Pi$. The first is diffusive (so $\dot L$ is a flux proportional to $\nabla\mu \sim \Pi/L\sim \sigma/L^2$, where $\sigma$ is unperturbed since $\lambda = 0$); $\zeta$ does not enter here. The second is hydrodynamic (balancing a viscous stress $\eta\nabla^2 v \sim \eta\dot L/L^2$ against $\nabla\!\cdot\! {\bf {\Sigma}} \sim \tilde\sigma/L^2$). 
When $\tilde\sigma = \sigma$, Eq.\ \eqref{nine} captures the well-studied crossover from Model B behavior, $L\propto (\sigma t)^{1/3}$, to viscous hydrodynamic (VH) coarsening, with $L\propto \tilde{\sigma}t/\eta$, at a length scale $L_\times \propto \eta^{1/2}$.  
Despite loss of TRS, a perturbative shift in $\tilde\sigma$ is unlikely to change this outcome, since Eq.\ \eqref{nine} rests on little more than dimensional analysis \cite{Bray}.

The opposite limit is that of pure MIPS, where there are no attractions between particles and every coefficient in Active Model H is set by far-from-equilibrium physics (primarily, the dependence of a particle's swim speed on the local density \cite{TC,sten1}). In a TRS system the form of $\mathcal{F}$ stems from a Hamiltonian that determines the diffusive (from $\nabla\mu$) and mechanical (from $\mathbf{\Sigma}$) currents in a thermodynamically consistent way. However, as already mentioned, in MIPS, $\mathcal{F}$ is merely a mathematical construct; there is no Hamiltonian (even for $\lambda = 0$) and hence no link between $\kappa$ and $\zeta$. Indeed, whereas $\kappa$ is always positive, $\zeta$ can have either sign as we now show.

The argument is based on simple kinematics as developed in \cite{CT}. For a system of swimmers with a propulsion speed $w(\rho)$ that depends on position (or particle density $\rho$) but not on swimming direction $\vec{\hat{u}}$, the first and second orientational moments of the distribution function $\Psi(\vec{r},\vec{\hat{u}})$ obey \cite{CT}
{\allowdisplaybreaks\begin{gather}%
\vec{P} = -\tau\nabla(w\rho) \;, \label{ten}\\
\mathbf{Q} = -\tau\frac{d-1}{2d}\nabla(w\vec{P})^{\mathrm{ST}} \;. 
\label{eleven}%
\end{gather}}%
Here, $\tau$ is the orientational relaxation time; the particle density $\rho$ is the zeroth orientational moment of $\Psi(\vec{r},\vec{\hat{u}})$; and ST denotes the symmetric traceless component. Note that $\rho$ and $\phi$ are related linearly; specifically if one sets $a=b=1$, one has $\phi = (2\rho-\rho_H-\rho_L)/(\rho_H-\rho_L)$, where $\rho_{L}$ and $\rho_{H}$ denote the low and high densities of coexisting phases, respectively \cite{NatCom}.
Equation \eqref{ten} is purely kinematic in origin: wherever $w\rho$ has a gradient, there are more particles pointing (i.e., moving) down this gradient than up it, causing nonzero $\vec{P}$ \cite{TCReview}. 

For self-propelled particles, the leading-order mechanical stress is, as for ALCs, caused by their exerting force dipoles on the fluid \cite{RMP}. In general we can write this as $\mathbf{\Sigma} = -\zeta_P\vec{P}\!\otimes\!\vec{P}/{\rho}-\zeta_Q\mathbf{Q}$ where $\zeta_{P}$ and $\zeta_{Q}$ are activity parameters. However, from Eq.\ \eqref{eleven} it follows that if $w$ is a function of density only (as assumed to this order in theories of MIPS \cite{sten1}), $\nabla\!\cdot\!\mathbf{Q}$ is a pure pressure gradient and thus ignorable. 
Expanding Eq.\ \eqref{ten} as $\vec{P} = -\tau (w\rho)'\nabla\rho$ with prime denoting a $\rho$ derivative, we recover Eq.\ \eqref{eight}, where
\begin{equation}
\zeta(\phi) = \frac{\zeta_P}{\rho} \left(\frac{\tau(w\rho)'}{\phi'}\right)^2 \;. \label{twelve}
\end{equation}
The sign of $\zeta$ then depends only on whether swimmers are extensile ($\zeta_P>0$) or contractile ($\zeta_P<0$). Time-reversal interchanges these two cases, so the active contribution to $\zeta$ in Eq.\ \eqref{eight} is odd under it, unlike any passive part.\footnote{Note that Eq.\ \eqref{eight} satisfies the general criterion for an admissible active stress presented in \cite{Scriven}.} Finally, although $\zeta_P$ vanishes for spherical swimmers at low density \cite{Ghose} it does not do so in general.\footnote{For an isolated isometric swimmer with orientation $\vec{\hat{u}}$ the only tensor from which to construct a stress contribution is $\vec{\hat{u}}\!\otimes\!\vec{\hat{u}}$, whose mesoscopic average at low but finite density gives $\mathbf{Q}$ but not $\vec{P}\!\otimes\!\vec{P}$ \cite{Rjoy2}. At high density, where swimmers interact strongly and correlations develop between orientations and interparticle coordinates, the reasons given in \cite{Rjoy2} for the vanishing of the $\vec{P}\!\otimes\!\vec{P}$ term no longer apply.} 

In keeping with our earlier discussion of Eq.\ \eqref{eight} we now suppress the $\phi$-dependence of $\zeta$. However, one could alternatively retain $\zeta(\phi)$ in Eq.\ \eqref{twelve} as part of a `best-fit' procedure to a more detailed kinetic theory of MIPS (see Appendix \ref{AIV}); we have checked numerically that the results are broadly similar to those with constant $\zeta$ reported in what follows.

Next, we assume that local diffusive relaxation normal to interfaces ensures that their local structure is only weakly perturbed by fluid motion, just as applies in passive Model H \cite{Bray}.
We thereby nonperturbatively recover Eq.\ \eqref{nine}, now with $\tilde\sigma = \zeta\sigma/\kappa$ (see Appendix \ref{AIII}). So long as swimmers are extensile, the two interfacial tensions in Eq.\ \eqref{nine} are positive, and the diffusive ($L\propto (\sigma t)^{1/3}$) and VH ($L\propto\tilde\sigma t/\eta$) regimes both remain intact. The prefactor of the VH coarsening is however shifted (possibly by a large amount), as is the crossover length $L_\times \propto (\eta\sigma/\tilde\sigma)^{1/2}$.

In contrast, very different physics now arises for contractile swimmers. Here, the interfacial tension $\sigma$ in the diffusive sector is positive as usual, but the mechanical one, $\tilde\sigma$, is negative. Equation \eqref{nine} still makes sense, but instead of a crossover from diffusive to VH coarsening there is a balance point $L_B\propto (\eta\sigma/|\tilde\sigma|)^{1/2}$ where diffusive coarsening is negated by VH anti-coarsening. Thus we predict that growth of domains will cease at this scale, to be replaced by a dynamic equilibrium where the diffusive shrinkage of  interfacial area is balanced by its production under the action of the contractile stress.
This stretching of interfaces can be understood as a combination of the kinematics -- causing swimmers at an interface to be aligned normal to it on average -- with the flow pattern around a contractile swimmer, which pulls fluid inwards axially and expels it around the equator. For a swimmer normal to the interface (with either polarity) the latter is equivalent to a negative interfacial tension (see Fig.\ \ref{fig1}). This argument also shows why any $\phi$-dependence of $\zeta$ (unless it changes sign) should not alter things qualitatively. Negative active tension has also been suggested in \cite{Bialke} by an argument that appears unrelated, in that it applies equally to contractile and extensile cases.

\begin{figure}[ht]
\centering
\includegraphics[width=0.7\linewidth]{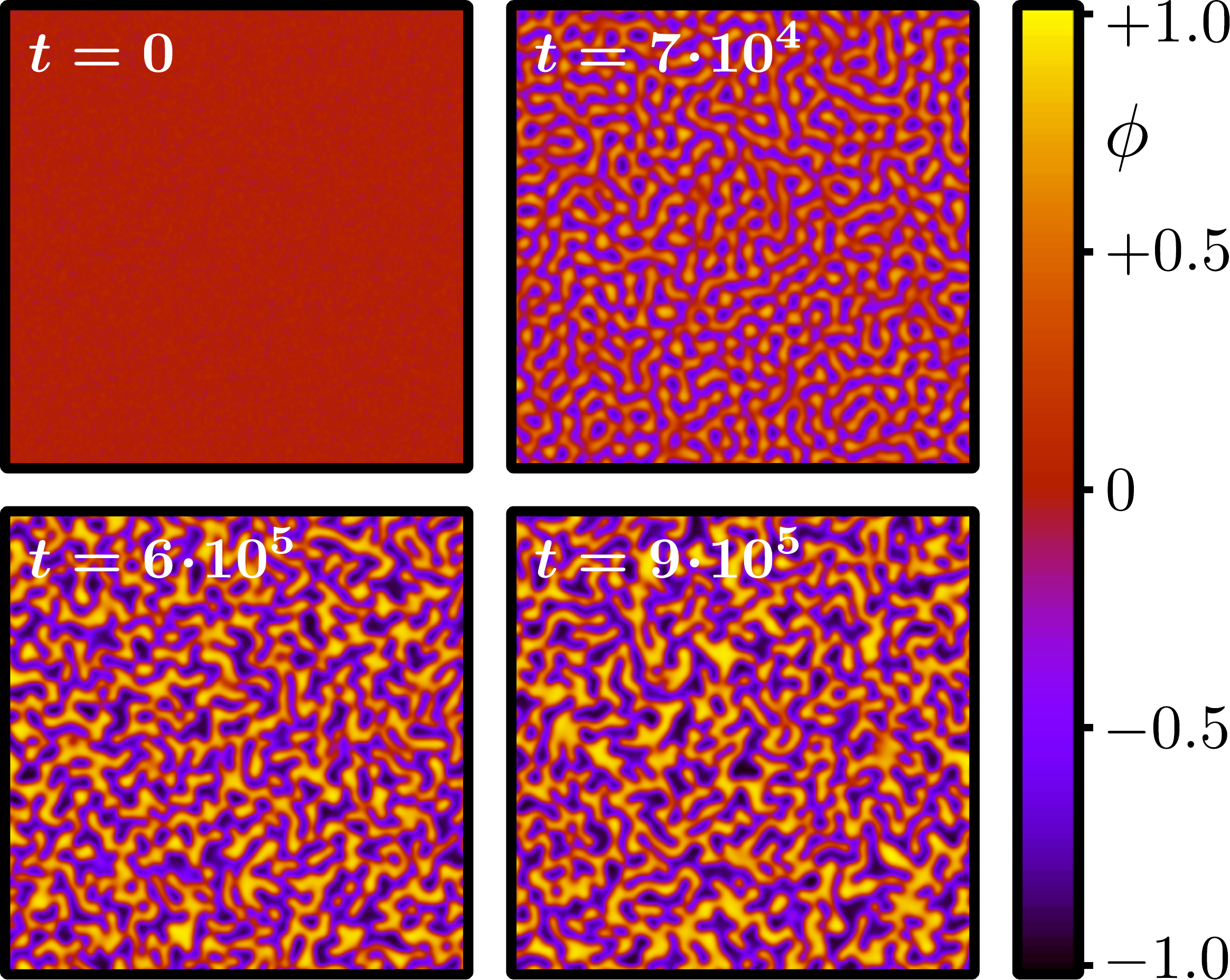}\vspace*{-1ex}%
\caption{Time series showing growth and arrest of domain structure in a two-dimensional contractile scalar active fluid with $\zeta=-0.001$ and box size $256\!\times\!256$.}
\label{fig3}%
\end{figure}

\textit{Numerical results:}
To test these ideas we have solved Active Model H numerically on a square lattice of size $l_x=l_y=256$ by using a noise-free (i.e., $\vec{\Lambda}=\vec{0}$)
hybrid Lattice Boltzmann (LB) scheme \cite{HLB}. We initialized the system in a mixed state with a small initial uniformly distributed noise with $-0.1<\phi(\vec{r},0)<0.1$. 
In our simulations we set $a=-0.004$, $b=0.004$, $\kappa=0.006$, $\lambda = 0$, and $\eta=1.67$, and varied $\zeta$. We checked that setting $\lambda\ne 0$ does not qualitatively affect our conclusions.  

These simulations confirm our expectation that contractile (but not extensile) activity should lead to the arrest of MIPS (see supplemental movies). This can be seen in Fig.\ \ref{fig2}, which shows $L(t)$ for active systems with various $\zeta$. Here, $L(t)$ is defined as usual through moments of the structure factor \cite{Kendon}. 
Snapshots of the dynamics corresponding to an arrested state are shown in Fig.\ \ref{fig3} and in supplemental movies; although the length scale is now fixed, the structure is highly dynamic. As $\zeta$ increases towards zero, the saturation length grows and for $-10^{-4}<\zeta<0$ coarsening does not arrest within the time window of our simulations. We speculate that it would do so eventually (at a length scale diverging as $\zeta \to 0^-$) although our data do not rule out a finite negative threshold above which arrest never occurs. 
The saturation length $L_B$ was argued above to obey $L_B\propto (\eta\sigma/|\tilde\sigma|)^{1/2}$. At fixed $\eta$ and $\kappa$ this gives $L_B\propto |\zeta|^{-1/2}$. Our data for $L_B(\zeta)$ instead approach a weaker power law at low activity and a plateau at larger values (see Fig.\ \ref{figSI} in Appendix \ref{AV}). The cause of this discrepancy is unclear, but might suggest that a force balance different from Eq.\ \eqref{nine} prevails in the highly dynamic states observed in strongly contractile systems.

The case $\zeta = 0$ equates to passive Model B (no coupling to fluid motion), whereas passive Model H is attained for $\zeta = \kappa$. The $L(t)$ curve is as expected in each case with final power laws close to 1/3 and 1, respectively.  
For passive Model H the crossover in $L(t)$ from early-time diffusion to late-time fluid flow is sigmoidal, and the negatively curved part has often been interpreted as a sublinear power law \cite{Kendon}. Whatever the precise interpretation is, extensile systems with $0<\zeta<\kappa$ interpolate smoothly between the two pseudo-passive limiting cases. Sigmoidal curves continue to be seen for systems with $\zeta > \kappa$, in which the tension in the fluid sector is higher than the one driving diffusion. The data does not exclude an eventual power that exceeds unity, but more likely reflects a prolonged sigmoidal crossover between diffusion and a linear growth with a large slope $\dot L \propto \tilde \sigma/\eta$. 

In summary, we have constructed Active Model H, a minimal model for scalar active matter coupled to a momentum-conserving solvent. TRS is broken by two effects. One, encoded by $\lambda$, causes shifts in the densities at which phases can coexist in diffusive equilibrium \cite{NatCom}. The second effect is new, and amounts to a mismatch between the interfacial tension $\tilde\sigma$ that drives fluid motion and the tension $\sigma$ that drives diffusive fluxes. For extensile particles (or weakly contractile ones in cases where phase separation is driven by attractive interactions rather than being purely motility-induced) both tensions are positive, and while their inequality violates TRS, its effects on coarsening dynamics are expected to be quantitative, not qualitative. The opposite is true for contractile particles which become aligned normal to the interface between phases, creating a flow pattern that stretches it ($\tilde\sigma<0$). Balanced by diffusion, this effect can cause the domain size $L(t)$ to saturate.

Finally, it is tempting to associate our prediction of arrest at a finite length scale $L_B$ in wet contractile scalar active fluids with observations of finite cluster formation, rather than full phase separation, in synthetic colloidal swimmers undergoing MIPS \cite{synthetics}. However, there are two objections to this. First, we do not know which, if any, of these systems are contractile. Second, most observations of cluster phases are in `nearly dry' systems: clusters are found within two-dimensional layers close to a momentum-absorbing boundary \cite{synthetics}. Nonetheless, our prediction is that arrested separation should be generic in contractile wet systems undergoing MIPS. We look forward to future experimental tests of this prediction.

\textit{Acknowledgments:} We thank Ronojoy Adhikari, Rosalind Allen, Sriram Ramaswamy, Joakim Stenhammar, and Julien Tailleur for useful discussions. This work was funded in part by EPSRC Grant EP/J007404. R.\,W.\ acknowledges financial support through a Postdoctoral Research Fellowship (WI 4170/1-2) and a Return Fellowship (WI 4170/2-1) from the Deutsche Forschungsgemeinschaft (DFG). M.\,E.\,C.\ is supported by the Royal Society.

\appendix

\section{\label{AI}Form of the diffusive current}
To second order in $\phi$ and third order in $\nabla$ the general form of the diffusive current $\vec{J}$ in Eq.\ \eqref{two} is 
\begin{equation}
\vec{J}=-\nabla\bigg(\Fdif{\mathcal{F}}{\phi}+\lambda_{0}(\nabla\phi)^{2}\bigg)+\epsilon\phi\nabla^{3}\phi + \vec{\Lambda}
+\mathcal{O}(\phi^{3},\nabla^{4})
\label{JF}%
\end{equation}
with a functional 
\begin{equation}
\mathcal{F}=\int \!\bigg(f(\phi)+\frac{\kappa(\phi)}{2}(\nabla\phi)^{2}\bigg)\dif^{d}r \;,
\end{equation}
where $f(\phi)$ and $\kappa(\phi)=\kappa_{0}+\kappa_{1}\phi$ are scalar functions, $\lambda_{0}$, $\epsilon$, $\kappa_{0}$, and $\kappa_{1}$ are constants, and $\vec{\Lambda}$ is -- in accordance with the fluctuation-dissipation theorem -- a zero-mean, unit-variance Gaussian white noise.
Equation \eqref{JF} is equivalent to 
\begin{equation}
\begin{split}
\vec{J}=&-\nabla\big(f'(\phi)-\kappa(\phi)\nabla^{2}\phi+\lambda(\nabla\phi)^{2}\big) \\
&+\epsilon\phi\nabla^{3}\phi
+\vec{\Lambda} +\mathcal{O}(\phi^{3},\nabla^{4})
\end{split}
\label{Ja}%
\end{equation}
with $\lambda=\lambda_{0}-\kappa_{1}/2$. 
We can rewrite Eq.\ \eqref{Ja} as  
\begin{equation}
\begin{split}
\vec{J}=-M(\phi)\Big(&\nabla\big(\tilde{f}'(\phi)-\tilde{\kappa}(\phi)\nabla^{2}\phi+\tilde{\lambda}(\phi)(\nabla\phi)^{2}\big) \\
&-\tilde{\lambda}'(\phi)\nabla\phi(\nabla\phi)^{2}\Big)
+\vec{\Lambda} +\mathcal{O}(\phi^{3},\nabla^{4})
\end{split}
\label{Jb}%
\end{equation}
with the density-dependent mobility 
\begin{equation}
M(\phi)=(\kappa(\phi)+\epsilon\phi)^{\epsilon/(\kappa_{1}+\epsilon)}
\end{equation}
and the scalar functions
{\allowdisplaybreaks\begin{align}%
\tilde{f}'(\phi)&=\int\!\frac{f''(\phi)}{M(\phi)}\,\dif\phi  \;, \\
\tilde{\kappa}(\phi)&=\frac{\kappa(\phi)+\epsilon\phi}{M(\phi)} \;, \\
\tilde{\lambda}(\phi)&=\frac{\lambda}{M(\phi)} \;.
\end{align}}%
When we approximate $M(\phi)$ in Eq.\ \eqref{Jb} by a constant and discard all contributions that are of third or higher order in $\phi$, we obtain
\begin{equation}
\vec{J}=-\nabla\big(f'(\phi)-\hat{\kappa}(\phi)\nabla^{2}\phi+\lambda(\nabla\phi)^{2}\big) +\vec{\Lambda} 
\label{Jc}%
\end{equation}
with $\hat{\kappa}(\phi)=\kappa_{0}+(\kappa_{1}+\epsilon)\phi$. 
Choosing $f(\phi)=a\phi^{2}/2+b\phi^{4}/4$ and approximating the function $\hat{\kappa}(\phi)$ by a constant $\kappa$ finally leads to Active Model B [see Eqs.\ \eqref{one}-\eqref{three}].
Notice that $\vec{\Lambda}$ is now in accordance with the fluctuation-dissipation theorem although the same does not apply to Eq.\ \eqref{Jb}.

\section{\label{AII}Form of the stress tensor}
To second order in $\phi$ and third order in $\nabla$ the divergence of a traceless and symmetric second-rank tensor $T_{ij}$ can always be written as
\begin{equation}
\partial_j T_{ij} = c_{0} \partial_j ( (\partial_i\phi)(\partial_j\phi) ) 
+ \partial_i h(\phi,(\nabla\phi)^{2},\nabla^{2}\phi)
\end{equation}
with a constant $c_{0}$ and a scalar function $h(\phi,(\nabla\phi)^{2},\nabla^{2}\phi)$ that depends on $\phi$, $(\nabla\phi)^{2}$, and $\nabla^{2}\phi$. 
This means that if we neglect multiples of the identity matrix, the stress tensor $\Sigma_{ij}$ can be written as
\begin{equation}
\Sigma_{ij} = c_{0} (\partial_i\phi)(\partial_j\phi) + \mathcal{O}(\phi^{3},\nabla^{3}) \;. 
\end{equation}

\section{\label{AIII}Dynamics of the domain size}
Here we present the derivation of Eq.\ \eqref{nine}, which describes the dynamics of the domain size $L(t)$, by standard power-counting arguments. 
For this purpose, we consider the domain size in the diffusive and in the hydrodynamic regime, but neglect the inertial regime.

In the diffusive regime, the dynamics of the domain size follows from the steady-state version of Eq.\ \eqref{five} together with Eq.\ \eqref{two}, where the noise $\vec{\Lambda}$ can be neglected: $\vec{v}\!\cdot\!\nabla\phi=\nabla^{2}\mu$.
To estimate the dynamics of $L(t)$, we replace $\vec{v}\to\dot{L}$, $\nabla\to 1/L$, and $\phi\to 1$. We further set $\lambda=0$, which allows to replace $\mu\to\sigma/L$, where $\sigma = (-8\kappa a^3/9b^2)^{1/2}$ is the interfacial tension in Model B \cite{Kendon}. These replacements lead to the dynamics in the diffusive regime: $\dot{L}\propto \sigma/L^{2}$.

In the hydrodynamic regime, in contrast, the dynamics of the domain size follows from the inertia-free version of Eq.\ \eqref{seven}: 
$\eta\nabla^2\vec{v} = \nabla p - \nabla\!\cdot\!\mathbf{\Sigma}$. We now replace $\vec{v}\to\dot{L}$, $\nabla\to 1/L$, and  
$p\delta_{ij}-\Sigma_{ij}\to\tilde{\sigma}/L$ \cite{Kendon}, where the tension $\tilde{\sigma}$ is different from $\sigma$ due to activity. 
This leads to the dynamics in the hydrodynamic regime: $\dot{L}\propto\tilde{\sigma}/\eta$.

The full dynamics of the domain size $L(t)$ is estimated by a linear interpolation between the diffusive and the hydrodynamic regimes, 
\begin{equation}
\dot{L}\simeq \alpha\sigma/L^{2} +\beta\tilde{\sigma}/\eta \;,
\label{eq:L}%
\end{equation}
where $\alpha$ and $\beta$ are dimensionless constants.
If we assume that local diffusive relaxation normal to interfaces ensures that their local structure is only weakly perturbed by fluid motion, as is the case in passive Model H \cite{Bray}, $\sigma$ in the first term on the right-hand side of Eq.\ \eqref{eq:L} is not affected by $\tilde{\sigma}$ in the second term deviating from $\sigma$. 
Since $\zeta$ from Eq.\ \eqref{eight} equals $\kappa$ and $\tilde{\sigma}$ in Eq.\ \eqref{eq:L} equals $\sigma$ in passive Model H, we can now approximate $\tilde{\sigma} = \zeta\sigma/\kappa$.

\section{\label{AIV}Explicit approximation of $\boldsymbol{\zeta(\phi)}$}
An estimate of $\zeta(\phi)$ is found by assuming a pure MIPS mechanism, and inferring from the Model H free energy an approximation for the density-dependent swim speed $w(\rho)$ (or equivalently $w(\phi)$) that appears in the microscopic expression for $\zeta(\phi)$ [see Eq.\ \eqref{twelve}].

First, we use the linear transform stated further above to relate the order parameter $\phi$ in Model H to the particle density $\rho$ in a system of self-propelled particles undergoing MIPS. This transform can be rewritten
\begin{equation}
\rho = \bar{\rho}+\phi\:\!\frac{\rho_{H}-\rho_{L}}{2} \label{eq:rho}
\end{equation}
with the central density $\bar{\rho}=(\rho_{H}+\rho_{L})/2$, where $\rho_{L}$ and $\rho_{H}$ are the low and high coexisting values of $\rho$ in the phase-separated state, respectively. When $\lambda = 0$, as considered here, these are found from a common tangent construction on the effective free-energy density \cite{TC,TCReview}
\begin{equation}
f_{\rho}=\rho(\ln(\rho)-1)+\int^{\rho}_{0} \!\!\!\ln(w(u))\,\dif u \;.
\label{eq:frho}%
\end{equation}
The density-dependent propulsion speed $w(\rho)$ of the active particles, which is needed to fully specify $\vec{P}$ and $\mathbf{Q}$, can be found in terms of $\phi$ by relating the two nonequilibrium chemical potentials $\mu_{\rho}=f'_{\rho}(\rho)=\ln(\rho)+\ln(w(\rho))$ and $\mu_{0}=f'_{0}(\phi)=a\phi+b\phi^{3}$, respectively expressed in the $\rho$ and $\phi$ representations. 
In the following, we choose $a=-b=-1$ for clarity, but the calculations can straightforwardly be extended towards general constants $a$ and $b$. 
Since $\rho$ and $\phi$ are related linearly, so should be their chemical potentials: we write $\mu_{\rho}=-\tilde{a}\mu_{0}+\tilde{b}$ with constants $\tilde a$ and $\tilde b$.\footnote{When $\mathcal{F}$ is the free-energy functional corresponding to the chemical potential $\mu_{\rho}=\delta\mathcal{F}[\rho]/\delta\rho$ and $\rho=\gamma_{0}+\gamma_{1}\phi$ and $\phi$ are related linearly, the chemical potentials $\mu_{\rho}=\delta\mathcal{F}[\rho]/\delta\rho=(1/\gamma_{1})\delta\mathcal{F}[\phi]/\delta\phi=(1/\gamma_{1})\mu_{\phi}$ and 
$\mu_{\phi}=\delta\mathcal{F}[\phi]/\delta\phi$ are proportional. Taking into account that chemical potentials are defined only up to a constant offset then yields a linear relation between $\mu_{\rho}$ and $\mu_{\phi}$.} 
This relation is equivalent to  
\begin{equation}
w(\rho)=\bar{\rho}\:\! w(\bar{\rho})\:\!\frac{e^{-\tilde a\mu_{0}}}{\rho} 
\label{eq:vrho}%
\end{equation}
with $\bar{\rho}\:\! w(\bar{\rho})=\exp(\tilde b)$, where the constant $w(\bar{\rho})$ can be estimated directly, or absorbed into $\zeta_P$ in Eq.\ \eqref{twelve}.

The constant $\tilde a$ should be chosen to match as closely as possible the bulk chemical potential contribution $\mu_{0}=-\phi+\phi^{3}$ in the $\phi^{4}$ theory to the non-polynomial form found from Eq.\ \eqref{eq:frho} via Eq.\ \eqref{eq:rho}.
This matching cannot be perfect and the best-fit criterion is therefore somewhat subjective. We choose to match the logarithmic derivative of the swim speed at the central density $\bar{\rho}$ of the phase coexistence,
$\dif\ln(w)/\dif\ln(\rho)\rvert_{\rho=\bar{\rho}}=\bar{\rho}w'(\bar{\rho})/w(\bar{\rho})$. 
The constant $\tilde{a}$ can then be expressed as
\begin{align}%
\tilde{a}=\frac{\rho_{H}-\rho_{L}}{2\bar{\rho}}\bigg(1+\frac{\dif\ln(w)}{\dif\ln(\rho)}\bigg\rvert_{\rho=\bar{\rho}}\bigg) \,. 
\label{eq:a}%
\end{align}%
This completes the specification of $w(\rho)$. The result can be used in Eq.\ \eqref{twelve} if one wants to replace the simplest version of Active Model H (which has constant $\zeta$) with an estimated $\zeta(\phi)$ found by assuming that the quartic effective free energy arises solely from a MIPS mechanism. 

If preferred, Eq.\ \eqref{eq:vrho} can be converted into a $\phi$-dependent propulsion speed $w(\phi)$ using the linear transform \eqref{eq:rho}:
\begin{equation}%
w(\phi)=\frac{2\bar{\rho}\, w(0)e^{-\tilde{a}\mu_{0}}}{2\bar{\rho}+\phi(\rho_{H}-\rho_{L})} \;. 
\label{vfinal}%
\end{equation}%

\section{\label{AV}Saturation length}
\begin{figure}[ht]
\begin{center}%
\includegraphics[width=\linewidth]{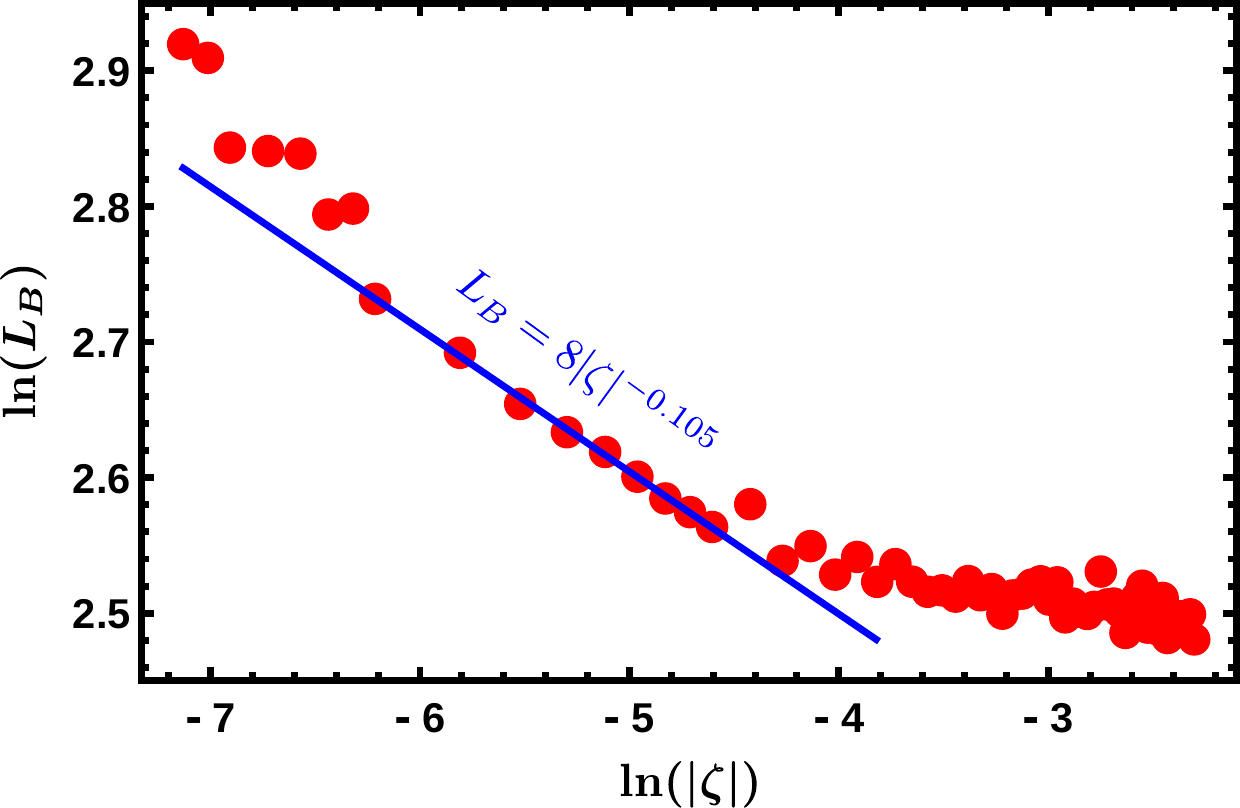}%
\caption{Lattice Boltzmann simulation data (red disks) and power-law fit curve (blue straight line) for the saturation length $L_{B}$ as a function of $|\zeta|$ for contractile activity ($\zeta<0$).}
\label{figSI}%
\end{center}%
\end{figure}

Figure \ref{figSI} shows the saturation length $L_{B}(\zeta)$ for contractile active particles (i.e., $\zeta<0$).
The data from our Lattice Boltzmann simulations show a weak power law $L_{B}=8 |\zeta|^{-0.105}$ at low activity ($0.002<|\zeta|<0.01$) and a plateau for larger activity ($0.03<|\zeta|$).
The exponent $\approx-0.1$ of the power law is larger than predicted by our simple theory, which yields an exponent of $-0.5$.

\vfill

\end{document}